# MediaWise – Designing a Smart Media Cloud

Dimitrios Georgakopoulos, Rajiv Ranjan, Karan Mitra, Xiangmin Zhou

*CSIRO ICT Centre*
*Canberra Australia*
{dimitrios.georgakopoulos,rajiv.ranjan,karan.mitra,xiangmin.zhou}@csiro.au

*Abstract*— The MediaWise project aims to expand the scope of existing media delivery systems with novel cloud, personalization and collaboration capabilities that can serve the needs of more users, communities, and businesses. The project develops a MediaWise Cloud platform that supports do-it-yourself creation, search, management, and consumption of multimedia content. The MediaWise Cloud supports pay-as-you-go models and elasticity that are similar to those offered by commercially available cloud services. However, unlike existing commercial CDN services providers such as Limelight Networks and Akamai the MediaWise Cloud require no ownerships of computing infrastructure and instead rely on the public Internet and public cloud services (e.g., commercial cloud storage to store its content). In addition to integrating such public cloud services into a public cloud-based Content Delivery Network, the MediaWise Cloud also provides advanced Quality of Service (QoS) management as required for the delivery of streamed and interactive high resolution multimedia content. In this paper, we give a brief overview of MediaWise Cloud architecture and present a comprehensive discussion on research objectives related to its service components.  Finally, we also compare the features supported by the existing CDN services against the envisioned objectives of MediaWise Cloud.

1. **Introduction**

The concept of a Content Delivery Network (CDN) [1] was conceived in the early days of the Internet but it took until the end of the 1990's before CDNs from Akamai and other commercial providers managed to deliver web content (i.e., web pages, text, graphics, URLs and scripts) anywhere in the world and at the same time meet the high availability and quality expected by their end users. Commercial CDNs achieved this by deploying a private collection of servers and a distributed CDN software system in multiple data centres around the world.

A different variant of CDN technology appeared in the mid 2000's to support the streaming of hundreds of high definition channels to paid customers. These CDNs had to deal with more stringent Quality of Service (QoS) requirements that were due to both the data scale and quality of experience required to delivery high definition video. This required active management of the underlying network and the use of specialized set-top boxes that included video recorders (providing stop/resume and record/playback functionality) and hardware decoders (e.g., providing mpeg 4 video compression/decompression). Major video CDNs where developed by telecommunications companies that owned the required network and had Operation Support Systems (OSSs) to manage the network QoS as required by the CDN to preserve the integrity of the high definition video content. Just like the original CDNs, video CDN also utilize a private collection of servers distributed around the network of the video service provider. The first notable CDNs in this category include Verizon's FiOS and AT&T's U-verse.

A more recent variant of video CDNs involves the caching video content in cloud storage and distribution of such content using third party network services that are designed to meet the QoS requirements of caching and streaming high definition video. Netflix's video CDN has been developed on top of Amazon AWS. CloudFront is Amazon's own CDN that uses Amazon AWS and provides streaming video services using Microsoft Xboxes. While Cloud-based [2][3] CDNs have made a remarkable progress in the past five years, they are still limited in the following aspects:

- CDN service providers either own all the services they use to run their CDN services or they outsource this to a single cloud provider. A specialized legal and technical relationship is required to make the CDN work in the latter case.
- Video CDNs are not designed to manage content (e.g., find and play high definition movies). This is typically done by CDN applications.  For example, CDNs do not provide services that allow an individual to create a streaming music video service combining music videos from an existing content source on the  Internet (e.g., YouTube), his/her personal collection, and from live performances he/she attends using his/her smart phone to capture such content. This can only be done by an application managing where and when the CDN will deliver the video component of his/her music program.
- CDNs are designed for streaming staged content but do not perform well in situations where content is produced dynamically. This is typically the case when content is produced, managed and consumed in collaborative activities. For example, an art teacher may find and discuss movies from different film archives, the selected movies may be then edited by the students. Parts of them may be used in producing new movies that will be sent to the students' friends for comments and suggestions. Current CDNs do not support such collaborative activities that involve dynamic content creation.
- CDNs do not support content personalization.

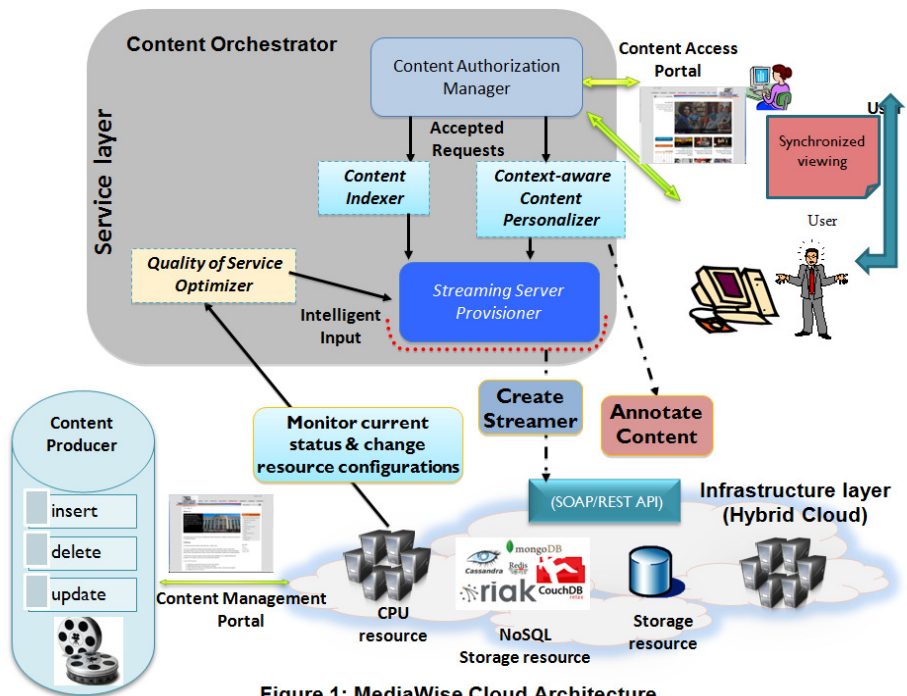

Figure 1: MediaWise Cloud Architecture

In this paper, we introduce the MediaWise Cloud platform and discuss its architecture. The MediaWise Cloud aims to address shortcomings of the current CDN technologies. In particular, Specifically, MediaWise Cloud supports the following novel capabilities:

- Exclusively utilize public Internet and public cloud infrastructure. Unlike CDNs, the MediaWise Cloud does not require the ownership of private networks and servers. Instead, it can utilize the cloud storage and compute services from virtually any public cloud provider. This provides additional flexibility for meeting QoS requirements (e.g., by staging content in public cloud storage "closer" to its consumers and by choosing the most cost-effective combination of public cloud service providers to deliver short term and long term content delivery services). To deal with the increasing complexity of QoS management, the MediaWise Cloud will monitor and learn the QoS-related performance of virtually all available public cloud services, and using this information for on demand prediction of expected QoS for media delivery requests.
- Supply content management services that can be used by content managers (a class of users that is currently not supported by CDNs) to manage content composition and delivery workflows for their end users.
- Support dynamic content delivery to enable collaborative activities.
- Provide seamless and personalized end-user experience. To achieve this goal, MediaWise will develop sophisticated user contexts for location-dependent media-related activities and user preferences.

The rest of this paper is organised as follows: Section 2 presents the main MediaWise innovations and gives examples of MediaWise applications. Section 3 describes the MediaWise Cloud Architecture and associated research challenges. Related work is reviewed in Section 3. Conclusion and future work are presented last in Section4.

2. **MediaWise Cloud Architecture & Research Objectives**

In MediaWise project, we are developing a cloud-based generic and scalable software framework called MediaWise Cloud Content Orchestrator for the production, consumption, and delivery of contents via public Internet. In rest of the paper, we refer to the terms MediaWise Cloud Content Orchestrator [4] and MediaWise Cloud interchangeably. The orchestrator will exploit hybrid clouds to offload computing, storage and content distribution functionalities in a cost effective manner. The high level architecture of MediaWise Cloud Content Orchestrator is shown in Figure 1. It is divided into two layers:

**Infrastructure layer:** This layer provides cloud-based hardware resources such as CPU, storage, routers and switches that are necessary to build a plug-and-play CDN. MediaWise project will consider aggregation of both public (e.g. Amazon EC2, Amazon S3, Ninefold, CloudCentral and GoGrid) and private (e.g., CSIRO cloud) cloud resources for hosting and delivering content. In general, cloud providers manage resources through hardware virtualization technologies [5] such as Xen, Citrix, KVM (open source), VMWare and Microsoft Hyer-V. Virtualization allows providers to get more out of hardware resources by allowing multiple instances of virtual resources to run at the same time. Each virtual resource believes it has its own hardware. Virtualization isolates the resources from each other, thereby making fault tolerant and isolated security behaviour possible.

**Service Layer:** This layer features a rich pool of services including content authorization manager, content indexer, context-aware content personalizer, streaming server provisioner and QoS modeller that MediaWise project researchers will design, implement and integrate for building Content Orchestrator. Notably, each of the aforementioned services has to perform certain orchestration operation on infrastructure level cloud resources, such as provisioning of a streaming server over an Amazon EC2 or indexing of contents over Amazon S3. Interaction with cloud resources can be done through Application Programming Interface (API) in specific programming languages such as Java, C#, Python and Ruby on Rails. Unfortunately, most of the existing APIs supported by cloud providers (e.g., Amazon, Microsoft Azure, GoGrid and Ninefold) are not compatible with each other. These providers tend to have their own proprietary APIs which are not explicitly designed for cross-cloud interoperability. To tackle such heterogeneities, there is a design requirement to enforce standardization across service implementations.

In this project, we will take advantage of recent developments in the context of standardized cloud APIs including Simple Cloud, Delta Cloud, JCloud, and Dasein Cloud, respectively. These APIs simplify the cloud programming task by implementing single API that abstracts multiple heterogeneous APIs exposed by cloud providers. Though aforementioned APIs can simplify the implementation of MediaWise content orchestrator that can work across multiple clouds, we still need to cater for the heterogeneities that prevail in terms of virtualization technologies, resource naming, and cost/performance.

## 2.1 Research Objectives

This section summarizes the research objectives of MediaWise Cloud. A comparative study of the existing CDN services against the research objectives of MediaWise is presented in Table I.

A. *Content creation and management*

Users and organizations involved in the creation of multimedia content play the role of content producers. The content they produced can be requested by and delivered to users playing the role of content consumers. Existing models of multimedia content creation are usually one-to-many (i.e., they involve one producer and many consumers) [14][15]. Content producers perform digital capture of multimedia data, and use software tools to perform multimedia post-production (e.g., to add graphics, audio, and titles to raw video clips), and finally produce a media product (e.g., a TV program, a Podcast, or a set of structured multimedia Web pages that link text, pictures, video and music content). The MediaWise project will focus its research on many-to-many models for creating novel forms of expressive and interactive content. In particular, as a starting point MediaWise will utilize existing Web 2.0 technologies for creating structured multimedia content (i.e., tools allowing users to create web pages linking video, pictures, audio, text, and graphics), but broaden existing Web 2.0 strengths with new technologies for creating novel forms of interactive multimedia content (e.g., as discussed earlier, MediaWise interactive content will involve virtual and augmented reality). In particular, novel research in the MediaWise project will focus on the following three content creation and management areas:

**Novel types of multimedia content.** Existing Web 2.0 technologies currently support the authoring of *structured multimedia content* (e.g., web pages linking images, sounds, videos, and animations). The MediaWise Project will extend and broaden existing Web 2.0 strengths with a new environment aimed at supporting the creation and consumption of interactive multimedia content (e.g., interactive audio and video), as well as other novel forms of multimedia content (e.g., *virtual* and *augmented reality*) that are currently not supported by existing Web 2.0 technologies and tools.

**Advanced content management services:** Multimedia content management includes the creation, adaptation, and composition of *content management* (CM) services. Basic CM services package multimedia content together with functionality that manipulates it. Examples of basic content services include functions for storing, indexing, and searching multimedia content [17]. CM services may be adapted for a specific purpose, or personalized for individual users or communities. Creation, adaptation, and composition of CM services are accomplished by (possibly collaborating) users that play the role of *content managers*. Complex CM services may be created by adapting or composing simpler CM services. For example, a basic content management service may index movies, TV programs and news for future search. A complex content management service may provide an interactive TV guide and its functionality may use the basic indexing service.

The MediaWise project will develop and demonstrate complex content management services for the learning, news and entertainment applications we discussed earlier. For example, the MediaWise project will build a software tool that allows users to collaboratively create non-linear multimedia presentations. For example, a teacher may create a new project. Teams of students can play the role of content creators and upload new content to the common project in the form of photos as well as audio and video recordings from mobile and desktop devices. Other students can become content managers and create a joint presentations by defining the relations (e.g., temporal, geographical, or topical) between these multimedia objects. Moreover we want to enhance Web 2.0 with novel technologies (for storing, indexing, searching, adapting, composing, and consuming streamed and interactive multimedia content) as well as innovative tools for collaborative editing, trends/hot topics/historical patterns identification and metadata feature management. The interactive video production will be further optimized by removing the redundant information that may introduced by multiple users. The redundant multimedia parts that are same in content are identified by incorporating the advanced near duplicate multimedia content detection techniques [16]. MediaWise research in collaborative multimedia content management will provide next generation CM services for these functions; as well develop technologies for automating the adaptation and composition of CM services. The MediaWise project will develop a novel environment for automating the adaptation and composition of CM services.

**Collaborative content management workflows**: The MediaWise project will develop novel technology for supporting collaborative workflows that manage the lifecycle (i.e., the creation, management, and consumption) of complex multimedia content. In particular, the MediaWise project will research a wide range of user collaboration styles, develop novel technology for collaborative workflows managing the orchestration of services and human activities for multimedia content creation, management, and consumption, and determine how to achieve greater collaboration scale, user participation, and efficiency. Collaborative workflows for multimedia content creation, management, and consumption may range from structured to unstructured. As an example of structured workflow that manages the lifecycle of multimedia content, consider the production of a web-based news program. The production of such a news program follows an established workflow process that involves the following steps: (1) capture and post-process new video clips of a newsworthy event, (2) search video archives for related video clips, (3) create/edit audio annotations for the new and possibly old video clips, (4) save the new video clips/audio annotations and enable future content-based search, (5) capture the video or audio presentation of a news script by a news anchor, (6) post all products on a website, and (7) enable both linear and non-liner viewing of the news products. At the other end of the spectrum, ad hoc multimedia content creation involves no explicit control flow or coordination between multimedia content producers or mangers. Consider an existing Web 2.0 service that allows users to post video clips of events they consider to be newsworthy. This involves unconstrained uploading and sharing of multimedia content. The MediaWise project will develop new technologies that accommodate these and any style of work or user preferences ranging from ad hoc to highly-structured activities for collaborative content creation and management. MediaWise will focus particularly on identifying, modelling, and automating collaboration patterns that increase the scale, user participation, efficiency, and automation of collaborative media creation and management activities.

B. *Ubiquitous content delivery*

Multimedia content can be subscribed to or requested by *content consumers*. The goal of multimedia content delivery is to provide a seamless multimedia experience to users. In its simplest form, this involves rendering the multimedia content to deal with the constraints of the network that carries the multimedia content and the characteristics of the specific device a user utilizes to receive the content and interact with it. Advanced multimedia delivery involves the development of *content delivery* (CD) services. CD services will interact with the network and appropriately adjust its QoS as needed to deliver specific multimedia content to a specific user. In particular, the MediaWise project will focus on research will leading to the development of CD services. When given a specific content and a specific content consumer, these CD services will adjust QoS characteristics based on the content requirements for maintaining its integrity, the device the user is using, his/her location, and the service contract. The MediaWise project will research content transformation to meet target device and network constraints. Another related research objective of MediaWise is to provide ubiquitous content delivery between producers/managers that may be distributed across many states or nations. The MediaWise project will develop technology for ubiquitous content delivery accommodating desktop users as well as mobile users who may meet in a coffee shop.

C. *Flexible Content Storage, Compression, and Indexing:*

Cloud storage resources allow content producers to store content on virtualized disks and access them anytime from any point on the Internet. These storage resources are different from the local storage (for example, the local hard drive) in each CPU resource (e.g., Amazon EC2 instance types), which is temporary or non-persistent and cannot be directly accessed by other instances of CPU resources. Multiple storage resource types are available for building Content Orchestrator. Naturally, the choice of a particular storage resource type stems from the format (e.g., structured vs. unstructured) of the content. For instance, Azure Blob [18] and Amazon S3 [19] storage resources can hold video, audio, photos, archived email messages, or anything else, and allow applications to store and access content in a very flexible way. In contrast, NoSQL (Not Only SQL) storage resources have recently emerged to complement traditional database systems [20]. They do not support for ACID transaction principles, rather offer weaker consistency properties, such as eventual consistency. Amazon SimpleDB [23], Microsoft Azure Table Storage, Google App Engine Datastore [24], MongoDB [25], and Cassandra are some of the popular offerings in this category. For example, in Amazon CloudFront [21], CDN contents are organized into distributions. A distribution specifies the location of the original version of contents. The distribution can be hosted on cloud storage resources such as Amazon S3 or Amazon EC2 CPU resources. With an increase in the scale and the size of content distribution, efficient indexing and storage become a critical issue. The challenge is further aggravated in case of live and interactive contents, where size of distribution (hence the indexing complexity) in not known in advance. Though cloud environments are decentralized by nature, existing application architecture tends to be designed based on centralized network models. To support efficient content production and consumption on scale of TeraBytes or PetaBytes, it is mandatory to design decentralized content indexing algorithm to enable access and search over large-scale database. It is worth noting that none of the existing cloud storage resources exposes content indexing APIs, it is up to the CDN application designer to come-up with efficient indexing structure that can scale to large content sizes. To help end-users find and retrieve relevant content effectively, and to facilitate new and better ways of media delivery using cloud resources, advanced distributed algorithms need to be developed for indexing, browsing, filtering, searching and updating the vast amount of information available in multimedia contents.

D. *Content personalization and contextualization*

We are just beginning to realize the power of location-aware services, especially for mobile devices, that aid users in their interaction with their immediate physical environment. Location-aware services allow users to become aware of physically proximate resources that they might not otherwise know about, provide convenience in finding and interacting with those resources, and enable interpersonal networking that takes physical location as well as ad-hoc community-building in account. Imagine young people planning their weekend meetings within a connected infrastructure which is location-aware: the group of young people meet ad-hoc, are aware of each other's locations and create media content using video cameras or chatting (through typing or directly through voice messages).

MediaWise will also support additional contextual information, including users' resources and capabilities for networking and computing, their work and leisure activities, their preferences, and the communities to which they belong. For the user, the benefits of content awareness, convenience, and community-building will be enabled by each such context. For the content producers, contexts can be viewed as a mechanism for mass customization that better meshes the needs and interests of each user with the multimedia capabilities at his/her disposal. The MediaWise project will support models of user's profiles which will store contextual information, including users' resources and capabilities for networking and computing, their work and leisure activities, their preferences, and the communities to which they belong. For the user, the benefits of content awareness, convenience, community searching and community building will be enabled by each such a user profile. For the content producers, this contextual information can be viewed as a mechanism for mass customization that better meshes the needs and interests of each user with the multimedia capabilities at his/her disposal. In addition, models of users' profiles will include the specification of competences, skills of users, and evaluation of users' former activities, *potentially in other communities*. Additionally the storage of profiles of content creators and managers is especially important to find users which fulfil the requirements of a given community.

Research objectives within the MediaWise project will include the following: (1) *Semi-automatic contextualization* – the automatic construction and maintenance of contexts of MediaWise users through mechanisms including user-modelling and data mining, but augmented with more interactive mechanisms such as knowledge elicitation. (2) *Personalization* – capturing the needs of users related to their evolving contexts so as to maximize the benefit to each user. (3) *Adaptation* – customizing content and services based on the context space so as to better meet the needs of the users who ultimately pay for them. (4) C*ommunity building* – which extends personalization to groups of users with various commonalities. (5) Finally, *cross-community aspects of users' profile* – modelling aspects of users' profiles which are relevant for various communities to which they belong, such as skills and competences, as well as aspects of users' profiles related through their activities in other communities, such as level of involvement.

E. *Community building*

The concept of community is at the core of the Web 2.0 [22]. In most applications of the Web 2.0, the structures of communities in terms of members' competences are pre-defined, e.g. in the Flickr.com application, a user may publish their pictures to predefined groups: family members, friend or others. Therefore, in current Web 2.0 applications, a user cannot tailor a community to his/her own needs, for instance defining an editor, a graphic designer and 5 journalists are required for a newspaper edition community. Nor can users easily find communities to which they may contribute, depending on their competencies, skills and centres of interest. Currently, users enter and leave communities in a chaotic way, depending mostly on recommendations of other users or links they may find while navigating.

The MediaWise project intends to provide users with tools for community-building, based on a new model of communities and users' profiles. The proposed model of communities will integrate the concepts of competences and skills to specify the requirements of the community to be built. The model of users' profiles will support, on the one hand, the specification of competences and skills of users and, on the other hand, evaluation of users' former activities, potentially in other communities. The latter information will be obtained via: (1) real time monitoring of events (e.g., initiation and completion of user activities) that create, manage, and consume multimedia content, and (2) tracing such events involving one or more users to user communities. Based on the models of communities and users' profiles, new algorithms matching community specifications with users' profiles will be proposed for 1) identification of communities of interest, and 2) creation of new communities. Finally, as the requirements of a given community may evolve through time, the proposed model of communities will support adaptation of community specifications during the lifetime of a given community. Similarly, the profile of a given user will evolve in time along with the evaluation of his/her former activities. Therefore, tools supporting community management will take into account the high dynamics of communities and users' profiles.

### F. *Quality of Service Optimizer*

It has been shown that one of the challenges in provisioning cloud resources for managing CDN application is uncertainty. Resource [26] uncertainty arises from a number of issues including user location, content type, malicious activities and heterogeneity. In some cases, media content delivery application may face with failure of resources or sometimes it may suffer from lack of sufficient resources. In this project, we will investigate the following issues for ensuring end-to-end QoS fulfilment in content production, consumption, and delivery.

**Optimizing cloud resource selection:** there are two layers in cloud computing [27] as shown in figure 1: a) service layer (e.g., Google App Engine, 3Tera Applogic, BitNami), where an administrator builds applications using APIs provided by the cloud and b) infrastructure layer (e.g., GoGrid, Amazon EC2), where an administrator runs applications inside CPU resources, using APIs provided by their chosen guest operating systems. Optimal CDN application performance demands bespoke resource configuration, yet no detailed, comprehensive cost, performance or feature comparison of cloud providers exists. This complicates the choice of cloud provider.

The diversity of offering at this layer leads to a practical question: how well does a cloud provider perform compared to the other providers? For example, how does a CDN application engineer compare the cost/performance features of CPU, storage, and network resources offered by Amazon EC2, Microsoft Azure, GoGrid, FelxiScale, TerreMark, and RackSpace. For instance, a low-end CPU resource of Microsoft Azure is 30% more expensive than the comparable Amazon EC2 CPU resource, but it can process CDN application workload twice as quickly. Similarly, a CDN application engineer may choose one provider for storage intensive applications and another for computation intensive CDN applications. Hence, there is need to develop novel decision making framework that can analyse existing cloud providers to help CDN service engineers in making selection decisions.

**Adapting to Dynamic CDN workload:** Media delivery CDN applications must accommodate highly transient, unpredictable content user's behaviours (arrival patterns, service time distributions, I/O system behaviours, user profile, network usage, etc.) and activities (streaming, searching, editing, and downloading). Yet many cloud providers contract service-level agreements which stipulate specific QoS targets, such as how fast a web page is served. It is therefore important to all parties that highly variable spikes in demand, caused by large number of simultaneous requests for a shared CDN service, do not degrade QoS [7][6].

It is critical that the content orchestrator is able to predict the demands and behaviours of hosted media applications, so that it can manage resources adaptively. Concrete prediction or forecasting models must be built before the demands and behaviours of a CDN application can be predicted accurately.

The hard challenge is to accurately identify and continuously learn the most important behaviours, and accurately compute statistical prediction functions based on the observed demands and behaviours such as request arrival pattern, service time distributions, I/O system behaviours, user profile, and network usage. The challenge is further aggravated by statistical correlation (such as stationary, short- and long-range dependence, and pseudo-periodicity) between different behaviours and activities of CDN application.

**Adapting to Uncertain cloud environment:** The availability, load, and throughput of hardware resources (CPU, storage, and network) can vary in unpredictable ways, so ensuring that CDN applications achieve QoS targets can be difficult. Worse still, hardware resource status can be changed intentionally through malicious external interference. The recent high-profile crash of Amazon EC2 cloud, which took down many applications, is a salient example of unpredictability in cloud environments. Theoretically, the elasticity provided by cloud computing can accommodate unexpected changes in capacity, failure, adding hardware resources when need, and reducing them during periods of low demand, but the decision to adjust capacity must be made frequently, automatically, and accurately to be cost effective.

### G. *Other research objectives*

The objective of the MediaWise project is to provide for *agility* in the creation, management, and consumption of content by small groups of users. For large groups of users and massive content, the MediaWise project will focus on providing maximum *efficiency*. Specifically, it should be noted that the MediaWise Project intends to specifically address scalability, reliability and flexibility issues. While scale may be defined in a variety of ways (e.g., by the number of end-to-end users involved in various content production, management and consumption roles, the number or sources and size of multimedia content, the group sizes of collaborating content producers and managers, the size of consumer communities, etc.), scale issues impact all functional areas and user roles as illustrated in figure 2. Reliability is achieved through the fault-tolerance feature of the system in the face of sudden load spikes such as flash crowds. Flexibility of the system can be realized through the ease of use, integration to existing system and on-demand deployment. We intend to test selected scenarios within a mobile city environment so that local, as well as mobile, scenarios can actually be explored.

### 3. **MediaWise Innovations and Sample Applications**

In the following sections we discuss three interactive applications that will be demonstrated and used in the rest of this proposal to help explain the planned innovations of the MediaWise project and illustrate its benefits. After discussing these applications, we will introduce the technical research areas of the MediaWise Project that constitute the major functions of the MediaWise cloud platform for these applications. Finally, we outline the innovations of the MediaWise project, categorized by technical research areas, using examples from the application areas we introduced earlier. Figure 2 shows the innovation areas of the MediaWise project as intersections of the project's research and functional areas, the roles users play in producing, managing, and consuming multimedia content, and related application areas.

H. *Applications and Impact areas*

In the following paragraphs, we describe the specific application areas we plan to use to demonstrate the impact of the MediaWise project. Having specific applications and developing demonstrations that illustrate the benefits of this technology will provide focus in our research and facilitate adoption of MediaWise.

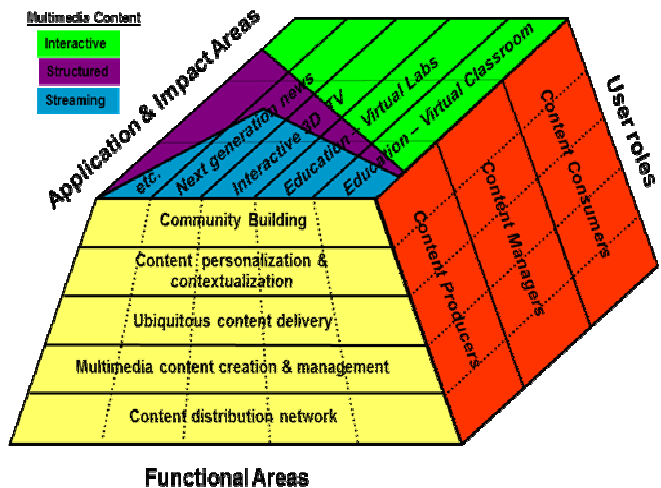

Figure 2: MediaWise Cloud research areas organized by functional areas, applications, and user roles

**Virtual classroom:** A virtual classroom is a crucial component of an e-learning system. It requires two main capabilities:

- *Synchronous communication and collaboration* for interactive teaching, questioning and answering, allowing class discussions, and supporting team work, and
- *asynchronous web-based knowledge management and dissemination* for making class-related material available to students, performing and submitting homework, and compiling with grading.

The MediaWise project will develop and demonstrate a next generation virtual classroom. Synchronous communication and collaboration will be supported by a high resolution (100M pixels) immersive videoconference environment that will be integrated with various state of the art tele-presence and shared workspace tools that are appropriate for remote teaching and class work. Provided workspace tools will include a virtual blackboard, screen sharing and co-browsing tools, as well as tools for scientific instrument sharing (e.g., virtual microscopy). Classroom tele-presense will be supported via low cost commercially available devices such as Microsoft's Kinect. This can achieve automatically tracking of the physical activities of all those in a virtual class room. For example, Kinect may be deployed in students' homes to track when two or more students raise their hands to ask or respond to a question. Automatic tracking such activities can be used to make the virtual classroom more responsive (e.g., notify the remote teacher that a student has a question or requests permission to speak) and to improve classroom fairness (e.g., by automatically queuing and responding to student requests in the order they are manifested by a physical activity).

Next generation asynchronous web-based knowledge management and dissemination will be provided via scientific wikis. Surprisingly, by looking at current practices in many fields of asynchronous web-based, collaborative, distributed knowledge production and dissemination (ranging from community-managed web repositories to complex software development), it is easy to observe that *innovative forms of scientific publications are still lagging behind*, and that the world of scientific publications has been largely oblivious to the advent of the Web and to advances in ICT. Scientific knowledge dissemination is still based on the traditional notion of "paper" publication. In this application area, we want to explore the MediaWise approach and how lessons learned from the social Web can be applied to provide a radical paradigm shift in the way scientific knowledge is created, disseminated, and maintained. We argue that novel technologies can enable a transition of "books" and "scientific publications" from its traditional "solid" form, (i.e. a crystallization in space and time of a scientific knowledge artifact) to a more fluid form (hereafter, what we call *Multimedia Scientific Knowledge Objects or MSKOs)*, that can take multiple shapes, evolves continuously in time, and is enriched by multiple sources. Conceptually, this application area is a multimedia counterpart of the traditional notion of scientific publications based on emerging and mature Web 2.0 services. We view the Multimedia Scientific Wiki service as a software application that manages MSKOs embodying a novel form of multimedia publications lifecycle. We consider MSKOs as being intrinsically:

- multi-media, i.e., they support different kinds of content, such as text, images, videos, slides, case studies, experimental datasets, and also include reviews and feedback by the community;
- multi-version, i.e., MSKOs and their constituents evolve over time as people contribute knowledge to them. They exist in multiple, incremental versions;
- multi-author, i.e., they enable the collaboration and contribution of a number of interested and expert authors on a specific MSKO, with different levels of "ownership" and control of the MSKO, and each able to claim credit and responsibility for the contribution; and
- multi-publication, i.e., they support the creation of new MSKOs by composing (and extending) existing ones.

Additional multimedia content in MSKOs includes video captures of demonstrations, experiments, and presentations, and other supplemental material that is essentially the complements of a publication, like experimental data, spatial and temporal information, etc., as well as the opinion and feedback of people in the community. All of these are part of the knowledge associated to an MSKO – and hence contribute to the creation of multimedia knowledge – as it facilitates the understanding of the authors' original contribution. MSKOs evolve in time as scientific knowledge progresses, and have many actors contributing to their creation and evolution, according to various lifecycle processes. Authors progressively add knowledge (delta-increments) to multimedia content. The community – including other teachers and students, progressively validates the quality of multimedia publications and adds value to it in the form of comments and feedback. Evolution of a publication is not conceived as a novel publication to be created, evaluated, and published anew (with significant loss of time for the community), but rather as the evolution of an existing MSKO, possibly by different authors, with each able to claim credit and responsibility for their contribution.

The paradigm we want to explore in this application is similar to what started to happen twenty years ago in software engineering with the progressive adoption of more agile and iterative development processes, from the spiral model to extreme programming up to "*social*", *open source development*. Furthermore, the open source and web communities teach us how to perform a "validation" and "credit attribution" of the work (which is key to people's careers and goals, and their continued participation) that is fair, relatively accurate, allows for high-quality artifacts to be generated, but is also lightweight (relative to peer review today) in terms of reviewing time requirements.

Besides supporting MSKOs and their lifecycle, the Multimedia Scientific Wiki service will involve the development of a prototype of a publication centre, i.e., a set of tools that manage the entire lifecycle of MSKOs and its continuous evolution.

**Virtual labs**: Today's Internet-based learning systems can be successfully applied to teach theoretical knowledge presented in the form of structured multimedia content. However, this form of presentation is often insufficient to teach practical skills. Serious games involving Virtual Labs provide users with interactive virtual reality environments, where they can collaboratively work on performing practical tasks and experiments. Serious games bring users exceptional freedom of experimentation. In a game environment, users can interact with virtual objects in a Virtual Lab similarly to the ways they would interact with real objects. For example, they can be confronted with interactive simulations of situations that they may not be able to experience in the real world. Another significant advantage is safety since unskilled learners are able to explore potentially dangerous situations without any risk of harm to themselves or damage to expensive equipment. Serious games may be particularly useful for presenting phenomena, which are:

- potentially dangerous (e.g., chemical or radioactivity experiments);
- macroscopic or microscopic (e.g., astronomical events and molecular movements);
- very fast or very slow (e.g., explosions and continental drift);
- normally hidden from view (e.g., inner workings of machines, human anatomy);
- normally inaccessible (e.g., a nuclear reactor, undersea life); or involve
- abstract concepts (e.g., magnetic fields, molecular forces).

There are many examples of subject domains where serious games have been used for educational purposes, e.g., geography, astronomy, chemistry, and physics. The level of success of applying serious games largely depends on the flexibility of learning environments to match particular user needs. The presented multimedia content can be tailored to the age, learning styles, and performance characteristics of users. Consequently, when users are provided with such highly-customized multimedia content, they become more interested and more engaged in the games and the degree of their satisfaction from the experience increases.

Serious games can serve as the focal point of vibrant eLearning communities. Multimedia content for learning may be created by content producers around the world that join together in communities working on a particular game. An example of a game community may consist of a user playing the role of an instructor, two advanced players, and several beginners. Within such a community two teams may be created. Each team is a sub-community, which must be composed of an advanced player, who is a team leader, and several beginners. The instructor may manage the way the game develops over time, and may judge the competitors and evaluate the achieved results.

In some serious games, player communities may be dynamic. Players may join or leave teams, and move from one team to another, if their profile fits requirements of a destination community. For example, a community may try to outbid another one for the best specialists by offering them better positions. The game provides additional interactive capabilities that help build gaming communities. When a player logs in, the game helps him/her find and re-join his/her team, or search for other gaming friends so he/she can create/join a new team.

There are several scenarios related to possible serious game environments. In one scenario the game may be immersed in a purely virtual environment, where multimedia content is presented in a 3D virtual space. In another scenario, the game may be located in an augmented reality environment, where multimedia content is displayed in real environments of the players. Augmented reality environments combine video streams presenting views of real environments with digital interactive multimedia content. Players using mobile devices could play an interesting variation of a serious game over physical terrain. In this case, the game environment includes real landmarks and real obstacles. A game may rely on searching for some real objects hidden in a real building or discovering some knowledge based on the information derived from the environment. This location-based variation of serious games fosters dynamic formation of location-based teams that are associated by physical proximity.

**Ubiquitous news:** There are several variations of this application. One variation revolves around a clearinghouse service for video and audio news. Creation of such multimedia news content would be performed by individuals that happen to be present in an event and use personal devices, such as video cameras and cell phones, for multimedia capture. This multimedia news content is consumed by users that search the new clearinghouse service for the news they need. Another variation of this news service may allow consumers to request news, even in situations where the event of interest is current but no multimedia news content is being collected or where there is no content because the news request is for a future event. To satisfy such consumer needs, the news service must obtain news content by communicating and negotiating with other users that may be willing to serve as news content creators. To address these diverse requirements, the news clearinghouse service indexes and stores old content, streams new content from current events, and facilitates collaboration between content creators and consumers to obtain new multimedia content. In one version of the news service, multimedia content is not managed in any way (i.e., is raw audio and video that is not combined with other content or adapted for its consumers). In a more traditional multimedia news service, that content will be managed (e.g., enhanced with titles and narrative, and adapted to fit a specific spot in the news program). Such content management may be performed by more specialized users that participate in another collaborative workflow that assembles a news program for consumers. When multimedia content is uploaded to the clearinghouse, contextual information (such as time and location) is automatically attached to the content. By using the clearinghouse the group is able to compose joint products, such as a "group trip report." These products also may be posted to a structured multimedia service that permits easy viewing by the family back home for multimedia content performed by more specialized users that participate in another collaborative workflow that assembles a news program for consumers. Another variation of ubiquitous news may involve location-based indexing and search capabilities for multimedia content. Such a service may allow a group of people (e.g., tourists or a school class on a field trip) to make movies and record narratives describing the points of interest.

Table 1: A comparative study of the existing CDN services against the research objectives of MediaWise

| CDN Application Provider | Content Creation and Management | Ubiquitous Content Delivery | Content Indexing | Content personalization and contextualization | Community Building | Quality of Service Optimization |
|---|---|---|---|---|---|---|
| Limelight Networks[8] | Browser based interface to upload static content; supports multiple media types; do not support dynamic content creation | Dependent on private Limilight networks backbone for content delivery; supports bitrate streaming as configured by content user; only limelight audio/video player supported | Title and Keyword based | No | No | Handled behind the scenes; content providers have no control over QoS; best effort QoS at network layer |
| Ooyala [9] | Browser-based interface to upload static content. Dynamic content creation is not supported. | Yes, to multiple devices using multiple formats | Title and Keyword based | Personalization is partially supported based on the type of devices to be used for media streaming | No | Yes, based on different CDN providers such as Akamai, |
| NetFlix [10] | N/A | Yes, to multiple devices using multiple formats | Title and Keyword based | Personalization is partially supported based on the type of devices to be used for media streaming | No | Yes, based on different CDN providers such as Akamai and now their own. |
| Akamai (Sola) and Ultraviolet [11] | Content can be managed partially, only for purchased content such as movies. | Yes, to multiple devices using multiple formats. | Title and Keyword based | Personalization is partially supported based on the type of devices to be used for media streaming | No | Yes, via Akamai. The content providers have no control over QoS provisioning. |
| MetaCDN [12] | Content cannot be created but can only be managed using pre-defined emplates | Yes, to multiple devices using multiple formats. | No | No | No | Yes but content providers have no control over QoS provisioning. |
| Rackspace [13] | Mainly content storage and provisioning | Yes, based on Akamai CDN | No | No | No | Yes, via Akamai. However, content providers have no control over QoS provisioning. |

## 4. Conclusions and Future Work

The growing ubiquity of the Internet and cloud computing is having significant impact on the media-related industries, which are using them as a medium to enable creation, search, management, and consumption of their contents online. We clearly articulate the architecture of MediaWise Cloud, its service components and associated research challenges, wherever applicable. Arguably, this paper is the first attempt at capturing the research and development challenges involved with engineering next-generation, *do-it-yourself CDN platform* using public cloud resources.

Our future work includes monitoring and learning the QoS-related performance of virtually all available public cloud services, and using this information for on demand prediction of expected QoS for media delivery requests. Other MediaWise innovations will include providing seamless and personalized user experience, and allowing users to collaborate in creating, managing, and consuming multimedia content virtually from anywhere and whichever means available to them. To achieve seamless and personalized user experience, MediaWise will develop sophisticated context-aware, location-dependent media-related activities and user preferences. It will develop technology that will present each user with media choices appropriate for his/her situation (i.e., location, task at hand, resources, skills, past activities, etc.). For seamless multimedia content experience, MediaWise will provide novel solutions for rendering such content to address network and user device constraints. To improve scale and efficiency of collaborative media creation and management activities, MediaWise will enhance user coordination by capturing and automating efficient collaboration patterns, activities, and processes. These will illustrate that the use of the MediaWise Cloud will help lower development costs and enable speeder development of applications; the MediaWise cloud will be used to develop and demonstrate three innovative applications in the areas of education, news and entertainment.